\documentclass{article}

\usepackage{arxiv}

\usepackage[utf8]{inputenc} % allow utf-8 input
\usepackage[T1]{fontenc}    % use 8-bit T1 fonts
\usepackage{hyperref}       % hyperlinks
\usepackage{url}            % simple URL typesetting
\usepackage{booktabs}       % professional-quality tables
\usepackage{amsfonts}       % blackboard math symbols
\usepackage{nicefrac}       % compact symbols for 1/2, etc.
\usepackage{microtype}      % microtypography
\usepackage{lipsum}		% Can be removed after putting your text content
\usepackage{graphicx}
\usepackage{doi}
\usepackage{booktabs}
\usepackage{multirow}
\usepackage{array}
\usepackage{siunitx}
\usepackage{moresize}

\graphicspath{ {./images} }

\title{Would you trust a vehicle merging into your lane? Subjective evaluation of negotiating behaviour in a congested merging scenario}

\author{ \href{https://orcid.org/0009-0003-5013-8008}{\includegraphics[scale=0.06]{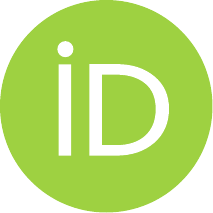}\hspace{1mm}Akinobu Goto}%\thanks{Use footnote for providing further
    \\
	Faculty of Engineering, Computer Science\\
	University of Bristol\\
	Woodland Road, Bristol, BS8 1UB, UK \\
    \\ Mobility and AI Laboratory \\
    Nissan Motor Co., Ltd.\\
    1-1 Morinosato-aoyama, Atsugi, Kanagawa, 243-0123, Japan \\
	\texttt{a\_goto@mail.nissan.co.jp} \\
	\And
	\href{https://orcid.org/0000-0001-9746-1409}{\includegraphics[scale=0.06]{orcid.pdf}\hspace{1mm}Kerstin Eder} \\
	Faculty of Engineering, Computer Science\\
	University of Bristol\\
	Woodland Road, Bristol, BS8 1UB, UK \\
	\texttt{kerstin.eder@bristol.ac.uk} \\
}

\date{}

\renewcommand{\headeright}{}
\renewcommand{\undertitle}{}
\renewcommand{\shorttitle}{Would you trust a vehicle merging into your lane?}

\newcommand\blfootnote[1]{%
  \begingroup
  \renewcommand\thefootnote{}\footnote{#1}%
  \addtocounter{footnote}{-1}%
  \endgroup
}

\begin{document}
\maketitle

\begin{abstract}
Aiming for a society where humans and automated vehicles can coexist cooperatively, understanding what constitutes cooperative and trustworthy behaviour is essential to designing automated vehicle controllers that enable the integration of highly automated vehicles into the real world. This study investigates how merging vehicles can gain trust from human-driven vehicles in a congested merging situation that requires explicit and implicit communication. 
Specifically, this study examines how the different behaviours of merging vehicles in the preparatory phase of the merge affect perceived trust from the perspective of the host vehicle in the mainstream lane. The findings suggest that transparent longitudinal positioning could improve the chance of successful merging, and cooperative deceleration during merging preparation could enhance the trust perceived by the host vehicle. Furthermore, the results reveal that, in time-sensitive situations where the merging vehicle approaches a lane closing point, prompt and decisive action of the merging vehicle encourages establishing trust with the host vehicle; any delay or hesitation can result in a lower level of trust. The results can provide valuable insights towards developing collaborative automated vehicles that improve safety and efficiency in real-world traffic situations that involve humans.
\end{abstract}

% keywords can be removed

\blfootnote{The copyright of this paper belongs to IEEE.}

\section{INTRODUCTION}
Automated driving technology has significantly progressed in recent years, focusing on preventing traffic accidents and improving comfort for drivers and passengers. To achieve a coexisting world between humans and automated vehicles (AVs, defined in~\cite{SAE}), it is important to gain the trust of other road users on shared roads. However, how to gain trust, especially in complicated driving scenarios that require interaction or negotiation among road users, is an open research question.

One challenging scenario is merging under congested traffic which requires cooperative driving~\cite{Zimmermann2}, which this study is focused on. When the main lane is heavily occupied and there is insufficient space to merge, the automated merging vehicle needs to negotiate with the host vehicle to make space. This demands cooperation and negotiation among road users, despite communication being highly restricted while driving~\cite{Chater}. To navigate such scenarios successfully, AVs have to engage in mutual interaction with human drivers. Explicit and implicit communication~\cite{Fuest} can be instrumental in fostering trust among road users~\cite{Kauffmann}. 

In order to develop an AV controller which can deal with such an interactive situation, an external perspective of how AVs are perceived by other road users would be important~\cite{Kauffmann}. Since human drivers can naturally handle such complicated scenarios, understanding human communication behaviour and preferences is an insightful starting point.

The cooperative driving scenario has attracted significant attention from researchers. However, there has been limited research focusing specifically on the communication behaviour of merging or lane-changing vehicles during the negotiation, regarding the subjective evaluation from an external perspective~\cite{Kauffmann}-\cite{Potzy1}. Hessen~\emph{et~al.} investigated the impact of activating the turn indicator~\cite{Hessen}. The timing of the turn indicator and the duration of lane change were examined by Kauffmann~\emph{et~al.}~\cite{Kauffmann}. 
In addition, Stoll ~\emph{et~al.} investigated the influence of situational factors such as the time to collision~\cite{Stoll}.
Besides, Potzy~\emph{et~al.} studied various parameters - including velocity, deceleration to the targeting position, lateral offset within the lane, the direction of lane change and the timing of the indicator - as communicating behaviours~\cite{Potzy1}. 

In spite of these valuable contributions, to the best of the authors' knowledge, no prior research has addressed the longitudinal positioning of the merging vehicle for effectively inducing cooperative behaviour from the host vehicle and enhancing trust. 
The relative longitudinal distance and velocity have been reported to be relevant to the perceived risk~\cite{Zhou}, and this perceived risk, in turn, has a correlation with the level of perceived trust towards an automated vehicle~\cite{He}.
Consequently, an erroneous positioning could potentially reduce the level of trust. Therefore longitudinal positioning should be considered an essential parameter in implicit communication and should be carefully designed.

In this paper, we commence by presenting a systematic view of cooperative merging. Subsequently, we conduct a subjective evaluation from the host vehicle's perspective considering various negotiating behaviour of the merging vehicle. Finally, we analyse and discuss the results. Our contributions can be summarised as follows:

\begin{itemize}
    \item We investigated the influence of the merging vehicle's longitudinal positioning during the negotiation process on the host vehicle's perceived trust.
    \item We found that the merging vehicle should consider its relative longitudinal positioning to the host vehicle and should employ a gentle deceleration rather than a harsh one during the negotiation to enhance trust.
    \item In urgent situations, it was found that investing too much time in communication reduces the level of trust.
\end{itemize}

\section{STUDY SETUP}
\subsection{Merging Phases}
\begin{figure*}[ht]
      \centering
      \includegraphics[width=1.0\textwidth]{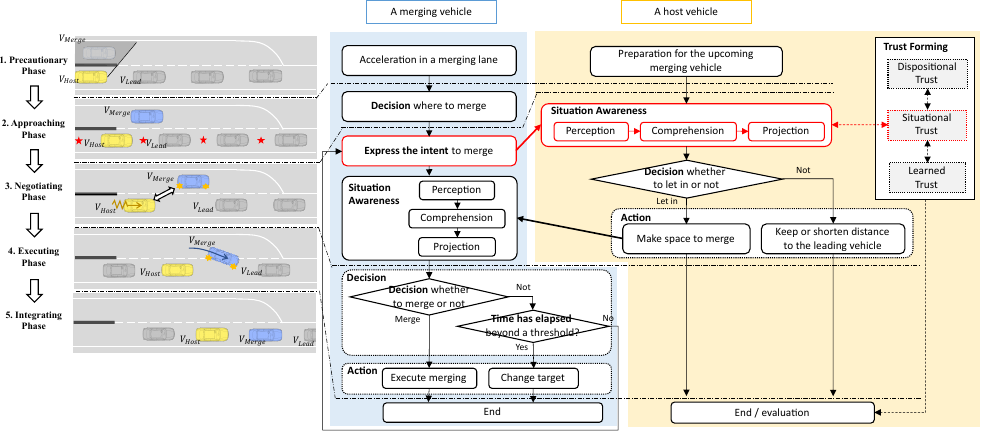}
      \caption{The five distinctive phases in the context of cooperative merging under congested traffic, corresponding communicative procedures for both a merging vehicle and a host vehicle, and the trust model. These consider the previous publication of Potzy~\cite{Potzy1} and Hoff~\cite{Hoff}.}
      \label{fig:diagram}
   \end{figure*}

To address this challenging driving scenario, we provide a systematic view of the entire cooperative merging comprising five distinctive phases, as illustrated in Fig.~\ref{fig:diagram}. In the first phase, a merging vehicle, $V_{Merge}$, approaches a merging section, without it being visible by the host vehicle, $V_{Host}$, running in a mainstream lane. In the next phase, the host vehicle can recognise the merging vehicle, and the merging vehicle selects a target space. The third is the negotiating phase, in which the merging vehicle shows its intent and encourages the host vehicle to make space. On the other hand, the host vehicle perceives the intent and decides whether to let it in. In the executing phase, the merging vehicle undertakes a lane change manoeuvre when the gap is sufficiently large, or it aborts merging into the initial target if the gap is insufficient. Finally, the host and the merging vehicle gradually adjust their longitudinal safety distance.

Among the aforementioned procedures, especially in the negotiation phase, the merging vehicle effectively expressing its intention to merge is crucial because it exerts a significant influence on subsequent behaviours of the host vehicle. Additionally, the situation awareness of the host vehicle could contribute to situational trust, one of the major factors in forming trust~\cite{Hoff}. Therefore, this study investigates how should a merging vehicle behave in the negotiating phase to gain trust while merging in congested traffic.
   
\subsection{Hypotheses}
\textbf{Transparent positioning}: While the empirical previous research using actual traffic data has analysed the speed synchronization characteristic of the merging vehicle to the leading vehicle~\cite{Wan}, it remains unclear where the merging vehicle should be positioned when adjusting the speed and indicating a turn signal to encourage cooperative behaviour from the host vehicle. According to the following reasons, it is postulated that a position of the merging vehicle closer to the leading vehicle would be more advantageous for the host vehicle than a position closer to the host vehicle. Firstly, detecting the traffic movement in front is more straightforward than the surrounding on the periphery~\cite{Green}. Secondly, the visual workload is associated with the frequency and duration of gaze changes~\cite{Pauzie}. Finally, system transparency helps build trust in automation~\cite{Hoff}. Therefore, a position close to the leading vehicle should be easily perceived by the host vehicle, for the host vehicle to gain trust.

%The second hypothesis concerns the impact of cooperativeness on the merging process. 
\textbf{Cooperative deceleration}: When the merging vehicle is positioned closer to the leading vehicle, then the merging vehicle will need to move backwards relative to the host vehicle to ensure a safety gap with respect to the leading vehicle. In this case, the way to decelerate can influence the level of trust the host vehicle perceives. Prior research has indicated that a driver's perception of risk when following and approaching a leading vehicle is significantly influenced by the time headway and the time to collision with the leading vehicle~\cite{Kondo}. Similarly, another study has demonstrated the applicability of the risk perception index even in the context of lane changing~\cite{Zhou}. Therefore, the merging vehicle should plan its deceleration manoeuvre, considering the host vehicle is located behind it in order to not force the host vehicle to cooperate.

%The third hypothesis is associated with the criticality of the situation. 
\textbf{Criticality of the situation}: In a merging scenario, the level of urgency is determined by the distance or time remaining until the end of the merging lane. A previous study on human driver expectations of AV's behaviour in a merging context ascertained that AVs should merge into smaller gaps in urgent situations compared to non-urgent ones~\cite{Potzy2}. Therefore, there is no need for the merging vehicle to carefully prepare for merging when approaching the lane closure point, as the driver of the host vehicle anticipates an efficient merging process.

%Hence, the merging vehicle should not insist on executing the prudent merging preparation behaviour when approaching the lane closing point, as the driver of the host vehicle anticipates an efficient merging process.

Based on these aspects, we hypothesise as follows: \textbf{(H1)} a merging vehicle should take a position closer to a leading vehicle rather than a host vehicle to be transparent (H1.1) and to enhance the level of trust (H1.2) when it starts indicating with a turn signal; \textbf{(H2)} a merging vehicle should plan its deceleration manoeuvre considering the host vehicle behind to be cooperative (H2.1) and to enhance the level of trust (H2.2); and that \textbf{(H3)} the closer to the lane endpoint, the more critical the situation, and therefore the more tolerant the host vehicle will be of the greater deceleration of the merging vehicle. This results in a higher level of reasonableness (H3.1) and trust (H3.2), in the critical situation, in spite of the merging intent not being expressed sufficiently clearly.

\section{METHODOLOGY}
To test the hypotheses, our study examines how the merging vehicle's behaviour is perceived from the host vehicle's perspective. 
As a starting point for the study, we set up a driving simulator test environment that ensures the repeatability of such complex scenarios.

%%%%%%%%%%%%%%%%%%%%%%%%%%%%%%%%%%%%%%%%%%%%%%%%%%%%%%%%%%%%%%%%%%%%%%%%%%%%%%%%
\subsection{Driving Simulator}
\begin{figure}[ht]
      \centering
      \includegraphics[width=0.48\textwidth]{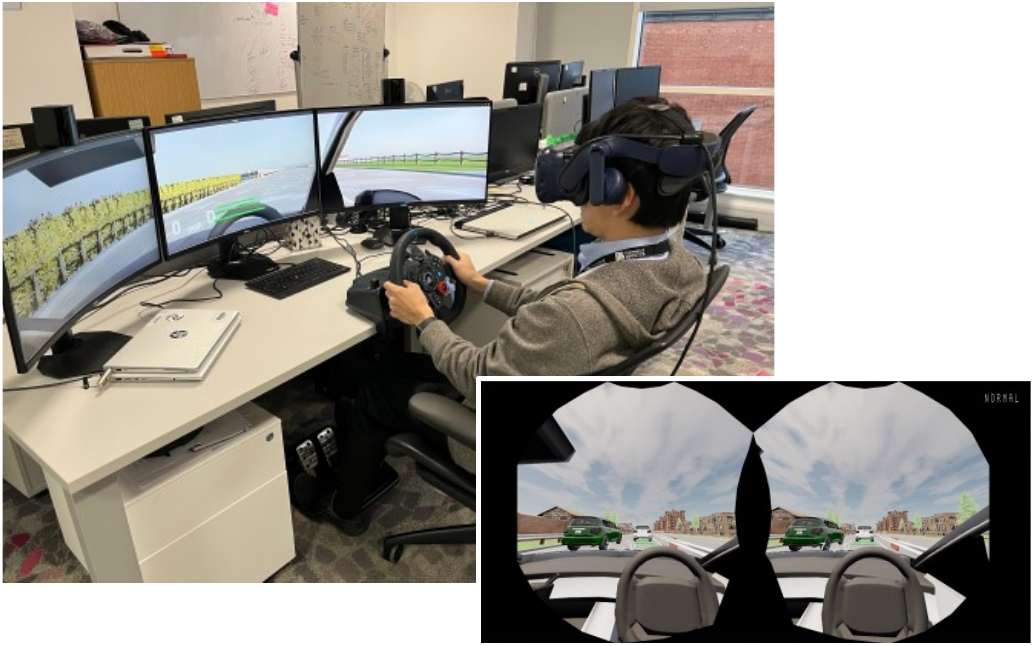}
      \caption{A participant driving the simulator with a VR headset and representative views of both eyes.}
      \label{fig:overview}
   \end{figure}

By using CARLA, an open-source simulator for developing automated driving technology that provides photo-realistic visualization capability~\cite{CARLA}, we built a driving simulator. We adopted virtual reality (VR) augmentation technology to realise an immersive feeling for participants based on the published leading literature by Silvera \emph{et~al.}~\cite{Silvera}, because it can be beneficial for participants to perceive the distance to other vehicles more precisely in an interactive driving scenario. Fig.~\ref{fig:overview} illustrates an overview of the driving simulator equipped with a Logitech G29 steering wheel and pedal controller that accepts driver inputs and HTC Vive Pro Eye as an image and audio generator.

\subsection{Driving Scenario and Parameters}
\begin{figure}[ht]
      \centering
      \includegraphics[width=0.48\textwidth]{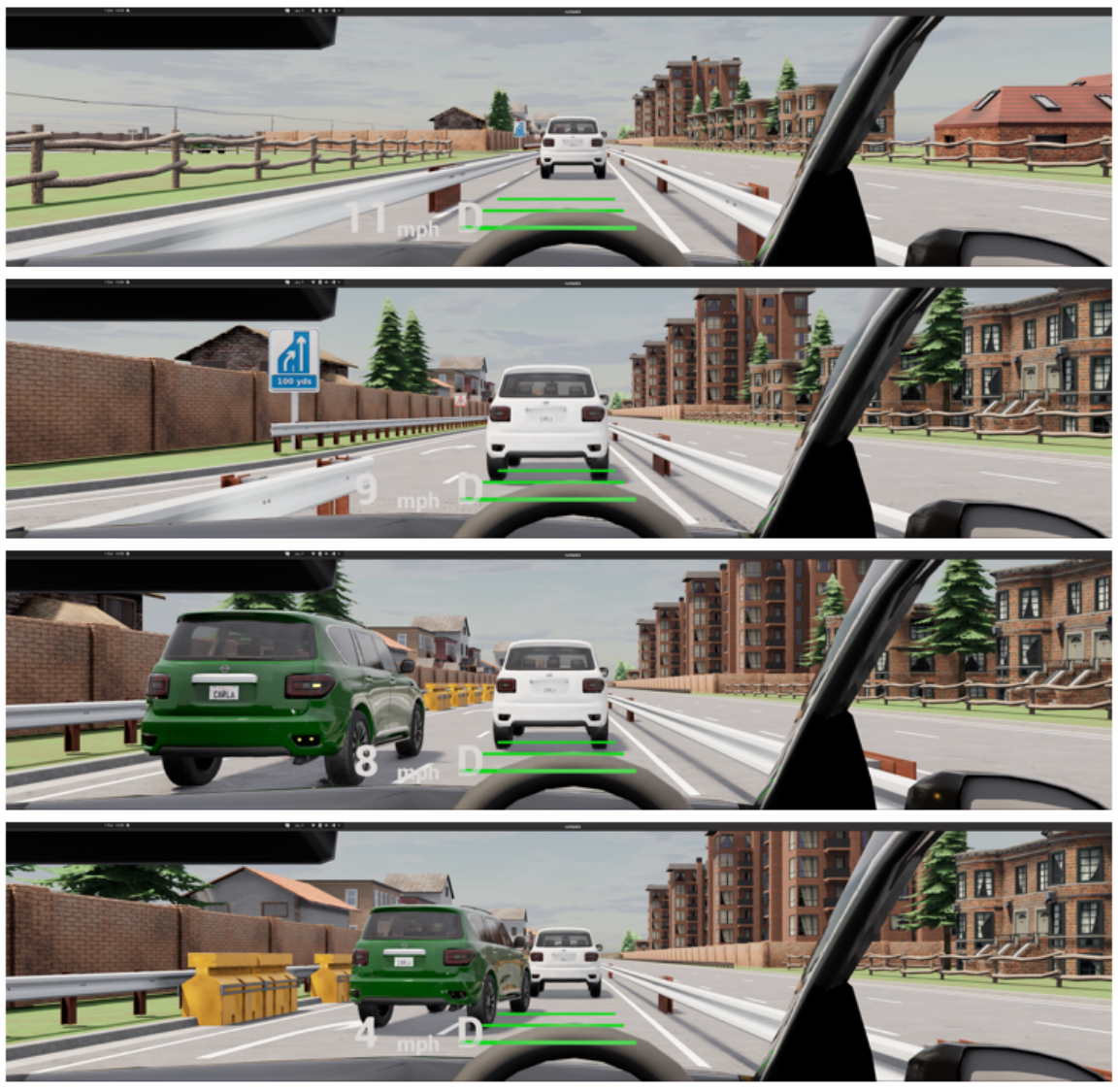}
      \caption{Sequence of the driving scenario operated in the experiment from the participant's point of view. }
      \label{fig:sequence}
   \end{figure}

Participants take control of the host vehicle in the simulation and evaluate the merging vehicle's behaviour from their point of view within a congested merging scenario. As illustrated in Fig.~\ref{fig:sequence}, the driving scenario unfolds as follows. Initially, the host vehicle, operated by a participant, follows the leading vehicle, which is running at 4.4~\unit{m/s} constantly, maintaining a gap of 10~\unit{m} displayed as green lines. After the host vehicle passes the beginning point of the merging section, the merging vehicle approaches the host vehicle from the left lane. After synchronizing its velocity to the leading vehicle, the merging vehicle starts indicating with its turn signal and negotiates using various behavioural patterns. Finally, it commences lateral movement if the host vehicle lets it in and if sufficient space to merge is available.
   
   \begin{figure}[ht]
      \centering
      \includegraphics[width=0.70\textwidth]{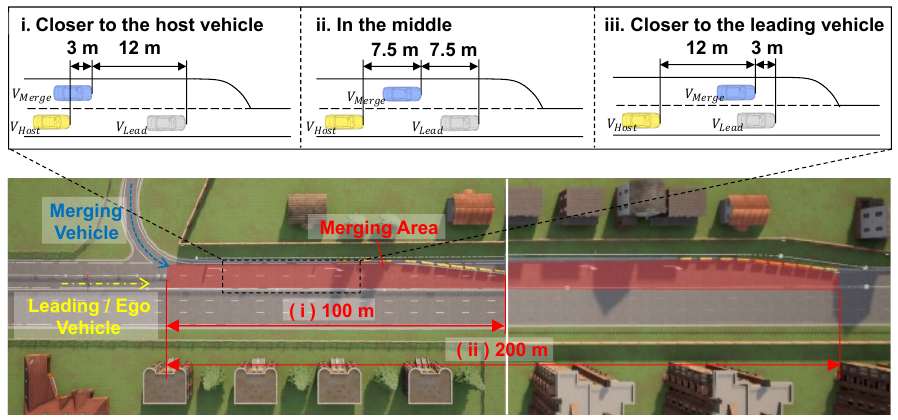}
      \caption{Dimensions of the scenarios. The upper figures show the relative position of each vehicle. The lower figures show the criticality of the situation; (i) a critical scenario, 100~\unit{m} in length from the beginning point of the merging section towards the end of the merging lane; and (ii) a non-critical scenario, 200~\unit{m} in length.}
      \label{fig:birds_eye}
   \end{figure}

\begin{table*}[htb]
    \begin{small}
    \centering
    \caption{Parameters in the experiment}
    \begin{tabular}{m{0.1cm} m{3.8cm} |
    >{\centering\arraybackslash}m{1.8cm} | % ID:1
    >{\centering\arraybackslash}m{1.8cm} | % ID:2
    >{\centering\arraybackslash}m{1.4cm} | % ID:3
    >{\centering\arraybackslash}m{1.4cm} | % ID:4
    >{\centering\arraybackslash}m{1.4cm} | % ID:5
    >{\centering\arraybackslash}m{1.4cm}}   % ID:6
    \toprule
        \multicolumn{2}{l|}{\multirow{2}{*}{\textbf{Parameters}} } & \multicolumn{6}{c}{\textbf{Scenario ID}}\\ 
        & & \textbf{1} & \textbf{2} & \textbf{3} & \textbf{4} & \textbf{5} & \textbf{6}\\ 
    \midrule
    \midrule
%        \multicolumn{2}{l|}{I. Transparency} & & & \multicolumn{4}{c}{}\\ 
%        & Longitudinal position of the merging vehicle when it starts the negotiation 
        \multicolumn{2}{l|}{\textbf{I. Transparency}} & & & \multicolumn{4}{c}{}\\ 
        & Longitudinal position of the merging vehicle when it starts the negotiation 
        & \textbf{i. Closer to the host vehicle} 
        & \textbf{ii. In the middle} 
        & \multicolumn{4}{c}{\textbf{iii. Closer to the leading vehicle}}\\ 
        \midrule
        \multicolumn{2}{l|}{\textbf{II. Cooperativeness}} &  &  & \multicolumn{2}{c|}{} & \multicolumn{2}{c}{} \\ 
        \multirow{2}{*}{} & Deceleration value of the merging vehicle
        & - 
        & - & \multicolumn{2}{c|}{\begin{tabular}{c}\textbf{i. Strong} deceleration \\ (-2.0 \unit{m/s^2}) \end{tabular}}
        & \multicolumn{2}{c}{\begin{tabular}{c}\textbf{ii. Weak} deceleration \\ (-0.5 \unit{m/s^2}) \end{tabular}} \\ 
        \midrule
        & Wait time to start deceleration after indicating turn signal
        & - 
        & - & \textbf{i. Short} wait-time (1.0~\unit{s}) & \textbf{ii. Long} wait-time (3.0~\unit{s})
        & \textbf{i. Short} wait-time (1.0~\unit{s}) & \textbf{ii. Long} wait-time (3.0~\unit{s}) \\ 
        \midrule
        \multicolumn{2}{l|}{\textbf{III. Criticality}} & \multicolumn{6}{c}{} \\ 
        & Distance between the beginning point of the merging section and the lane closing point
        & \multicolumn{6}{c}{\textbf{i. Close (100~\unit{m})  / ii. Far (200~\unit{m}) }} \\ 
    \bottomrule
    \end{tabular}
    \label{tab:parameters}
    \end{small}
\end{table*}

In the negotiating phase, the automated merging vehicle is programmed to use different parameters, as listed in Table~\ref{tab:parameters}. This leads to various behavioural patterns according to the three hypotheses. Firstly, the transparency parameter corresponds to longitudinal positioning, the specific dimensions of which are illustrated in Fig.~\ref{fig:birds_eye}. Secondly, the cooperativeness variants encompass deceleration and wait-time, which are designed to examine subjective differences in the evaluation of the merging vehicle's deceleration behaviour, particularly in establishing a safety gap to the leading vehicle. Therefore, once the safety gap is established, the merging vehicle accelerates and synchronises its speed with the leading vehicle. In other words, the deceleration is operated within the merging lane, which does not persist at the start of the merging execution. Particularly, scenarios ID-1 and ID-2 do not involve the cooperativeness parameters since they already have enough space between the merging vehicle and the leading vehicle. In these scenarios, the merging vehicle initiates the executing phase as soon as the host vehicle makes sufficient space to merge. Finally, the criticality parameter is the distance between the merging vehicle and the lane closing point, which is adjusted on the maps, as illustrated in Fig.~\ref{fig:birds_eye}. Lastly, there is a total number of 12 scenarios.

In the executing phase, the merging vehicle employs gap acceptance theory~\cite{Hidas}, to decide whether or not to merge, setting a minimum gap threshold of 2~\unit{m}. Additionally, the duration of executing the lane change is maintained at approximately 6~\unit{seconds}. These parameters remain constant across all driving scenarios. However, when the host vehicle fails to provide adequate space to let in the merging vehicle, the merging vehicle responds by decelerating near the lane closing point and subsequently merging behind the host vehicle once it has passed. 

In this study, the road layout was established by configuring the pre-installed Town06 asset in CARLA, and necessary modifications and additions were made to incorporate traffic signs, obstacles and road markings, announcing the merging section. Furthermore, to ensure the merging operation occurs within the lane, guardrails were installed along both sides of the driving lane to discourage unauthorised lane changes.

\subsection{Subjective Evaluation}
\begin{figure}[ht]
      \centering
      \includegraphics[width=0.6\textwidth]{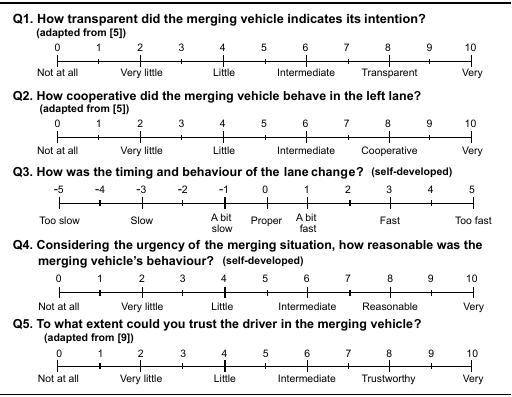}
      \caption{Questions and grades asked in the questionnaire}
      \label{fig:questionnaire}
   \end{figure}

Participants were asked to answer 5 different questions to evaluate each merging vehicle's behaviour after each scenario. As shown in Fig.~\ref{fig:questionnaire}, the items of the questionnaire consist of 1.\ transparency of the intent, 2.\ cooperativeness of the behaviour, 3.\ appropriateness of the merging timing, 4.\ reasonableness of the situation, and 5.\ trust, rated on a scale of 0 to 10 (although 3.\ the timing is rated from -5 to 5). Each question except 3 is related to each hypothesis of H1.1 to H3.2. 

\subsection{Experimental Procedure}
Each experiment took about 30~\unit{minutes}. Having passed an initial practice session to acclimatise to driving in the simulation environment, the participants experienced the 12 different driving scenarios in randomised order to reduce unconscious bias. The questionnaire was designed so that participants were able to answer the questions in the VR world using a motion controller without having to repeatedly put on and take off the VR headset, allowing them to focus on the evaluation.

%%%%%%%%%%%%%%%%%%%%%%%%%%%%%%%%%%%%%%%%%%%%%%%%%%%%%%%%%%%%%%%%%%%%%%%%%%%%%%%%
\subsection{Ethical consideration}
This experiment was approved by the Faculty of Engineering Research Ethics Committee of the University of Bristol (Ref: 11492). Considering the possibility to get motion sickness when using a VR environment, participants were told to be able to stop whenever they wanted. Prior to the experiment, we conducted the sample size estimation not to waste the resources of participants.

\section{RESULTS AND DISCUSSION}
25 Participants (16 males, 8 females, and 1 other) were recruited within the university community through an online recruitment tool or word of mouth. The age of participants ranged from 21 to 65 years (mean (M) = 37.75, standard deviation (SD) = 11.53). All had valid driving licences in the UK or EU. The driving experience of participants ranged from 3 to 47 years (M = 17.71, SD = 12.06) and annual mileage ranged from \numrange{3200}{56000}~\unit{km} (M = 12800, SD = 13485.97).

\subsection{Transparency: Longitudinal Position}

\begin{figure}[ht]
      \centering
      \includegraphics[width=0.6\textwidth]{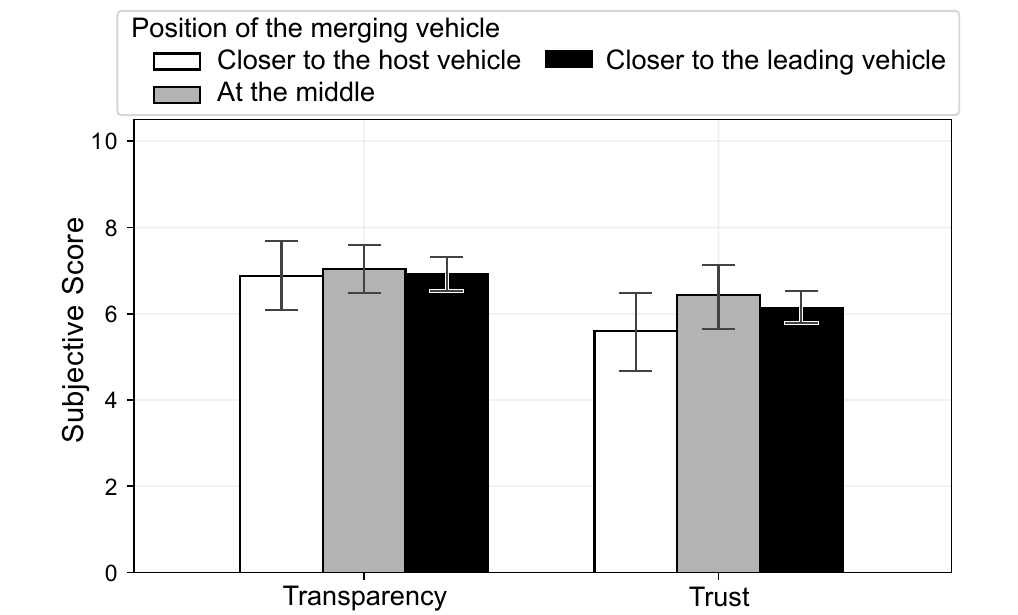}
      \caption{The subjective scores of transparency (H1.1) and trust (H1.2) according to the longitudinal position of the merging vehicle (mean and 95\% confidence interval). The results of one-way ANOVA show transparency: F = 0.121, p-value = 0.886 and trust: F = 1.042, p-value = 0.355.}
      \label{fig:result1.2}
   \end{figure}

Fig.~\ref{fig:result1.2} shows the outcomes of the subjective evaluation according to the first hypothesis with the different longitudinal positioning strategies. This graph specifically represents the results of the non-critical scenarios (200~\unit{m}), and the positioning close to the leading vehicle encompasses four patterns of different deceleration rates, which correspond to scenarios, ID-3 to ID-6. Upon examination of this graph, it becomes evident that the p-value resulting from the one-way analysis of variance (ANOVA) surpasses the threshold of 0.05, signifying a lack of statistically significant differences, across both aspects, transparency and trust.

\begin{figure}[ht]
      \centering
      \includegraphics[width=0.6\textwidth]{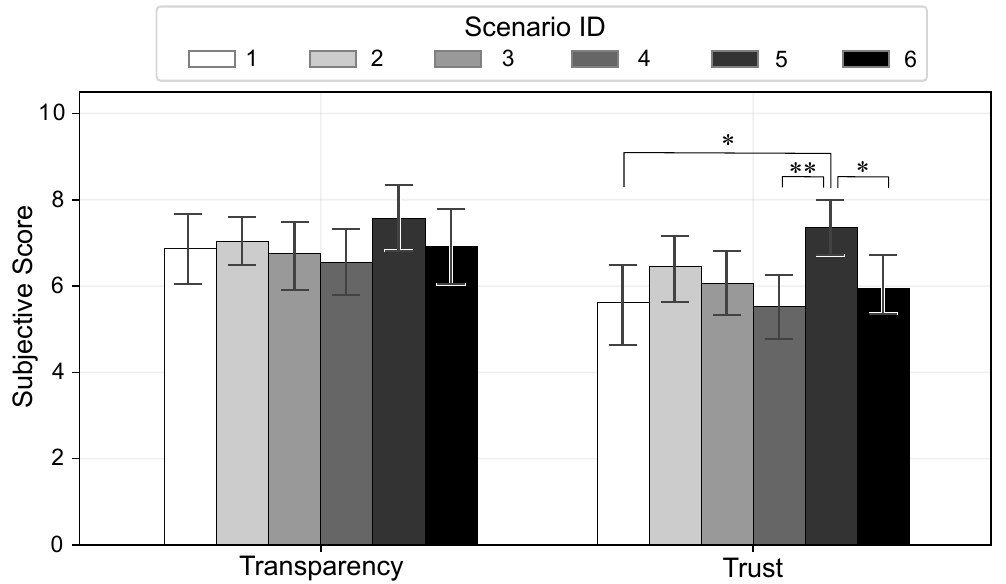}
      \caption{The subjective scores of transparency and trust according to the scenario IDs (mean and 95\% confidence interval). The results of one-way ANOVA show transparency: F = 0.548, p-value = 0.739 and trust: F = 2.625, p-value = 0.026.  Data shown as * p < 0.05, ** p < 0.01}
      \label{fig:result1.appendix}
   \end{figure}

On the other hand, the result of the subjective evaluation for each scenario ID, as seen in Fig.~\ref{fig:result1.appendix}, indicates that even if the vehicle is positioned close to the leading vehicle, without the proper subsequent deceleration the merging vehicle will not be able to gain trust. In fact, compared to scenario ID-1, close positioning to the host vehicle, scenario ID-5, close positioning to the leading vehicle, accomplished a significantly higher level of trust confirmed by the post hoc Tukey's test. However, in the other scenarios with the same position, ID-3, ID-4 and ID-6, no significant differences were observed compared to ID-1. Therefore, properly implementing deceleration actions, such as gentle and prompt deceleration we discuss in the following subsection, could serve as an essential incidental factor to the longitudinal positioning for enhancing trust.

\begin{figure}[ht]
      \centering
      \includegraphics[width=0.5\textwidth]{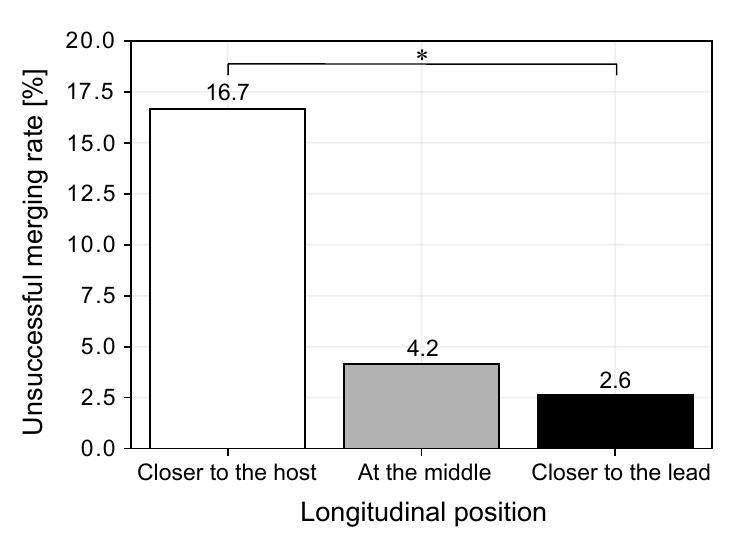}
      \caption{The unsuccessful merging rate for each longitudinal position of the merging vehicle, calculated individually for each longitudinal position. McNemar's test with * p < 0.05.}
      \label{fig:result1.1}
   \end{figure}

Additionally, we simultaneously confirmed the objective fact caused by the positioning of the merging vehicle. Specifically, as shown in Fig.~\ref{fig:result1.1}, McNemar's test verified that the risk of unsuccessful merging significantly increased when the merging vehicle was positioned closer to the host vehicle compared to the position closer to the leading vehicle. An unsuccessful merge is a situation where the original host vehicle refuses to give way to the attempting merging vehicle, resulting in merging behind it.

Based on the above results and discussion, this experiment did not provide conclusive evidence regarding the validity of the first hypothesis. However, it can be inferred that irrespective of the positioning, as long as the merging vehicle is clearly ahead, speed-synchronizing and activation of indicators communicate the intent to merge. Moreover, to encourage successful merging as much as possible, the merging vehicle should be situated closer to the leading vehicle than the host vehicle, or in the middle of those.

\subsection{Cooperativeness: Deceleration}

\begin{figure}[ht]
      \centering
      \includegraphics[width=0.6\textwidth]{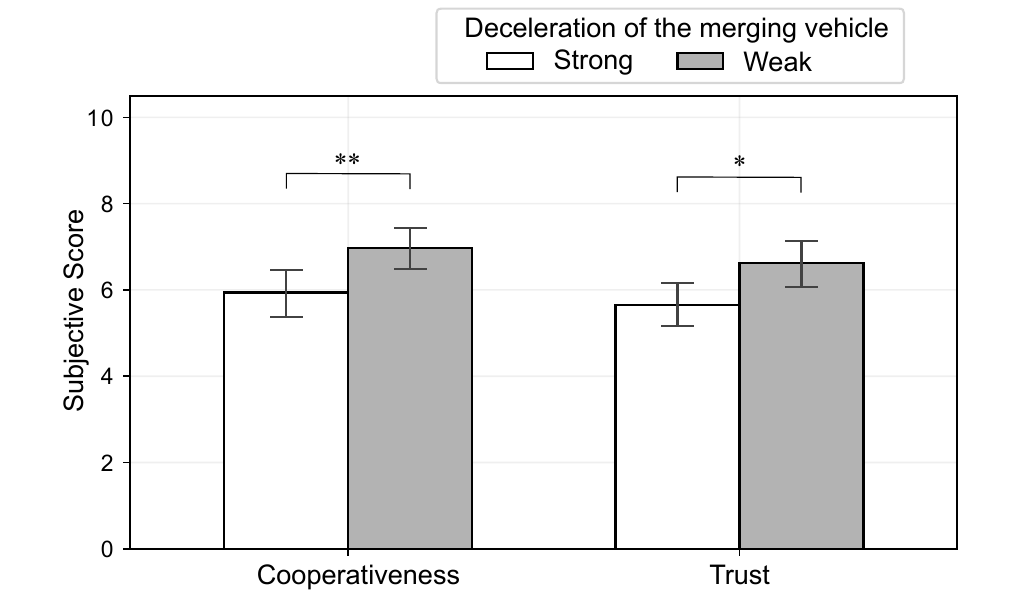}
      \caption{The subjective score of cooperativeness (H2.1) and trust (H2.2) according to the deceleration value of the merging vehicle (mean and 95\% confidence interval). The result of the paired t-test shows that cooperativeness: t(24) = -2.846, p-value = 0.005 and trust t(24) = -2.530, p-value = 0.013. Data shown as * p < 0.05, ** p < 0.01}
      \label{fig:result2.1}
   \end{figure}

The results for the second hypothesis are explained below. Fig.~\ref{fig:result2.1} presents the subjective evaluation outcomes for each deceleration scenario. It is evident that the weak deceleration scenario (-0.5~\unit{m/s^2}) produced significantly higher scores in terms of cooperativeness and trust compared to the strong deceleration scenario (-2.0~\unit{m/s^2}). Therefore, H2.1 and H2.2 are supported. Particularly, the strong deceleration scenarios encompassed scenarios ID-3 and ID-4, while the weak scenarios encompassed ID-5 and ID-6, respectively.

On the other hand, we consider the influence of different waiting times to start deceleration on the subjective evaluation. Fig.~\ref{fig:result1.appendix} illustrates the trust scores for each scenario ID on the right side. Among the scenarios characterised by weak deceleration (-0.5~\unit{m/s^2}), scenario ID-5 has a short wait-time of 1.0~\unit{s}, while ID-6 has a long wait-time of 3.0~\unit{s}. It shows that a prolonged wait-time leads to a significant decrease in the trust score compared to a shorter wait-time, even when the merging preparation operation is performed in a cooperative manner at a weak deceleration level.

These findings allow us to conclude that in order to foster cooperative behaviour and gain the host vehicle's trust, the merging vehicle ought to decelerate weakly (-0.5~\unit{m/s^2}) within a 1-second time window when establishing a safety gap with the leading vehicle.

\subsection{Criticality: Distance to the End of Merging Lane}

\begin{figure}[ht]
      \centering
      \includegraphics[width=0.6\textwidth]{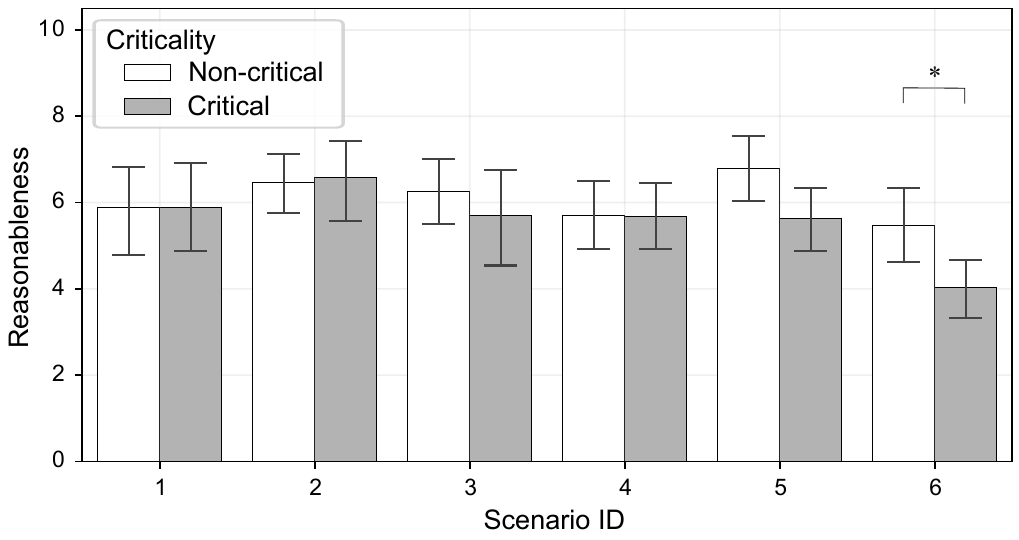}
      \caption{The reasonableness score (H3.1) according to the criticality of the situation for each scenario ID (mean and 95\% confidence interval). The result of the paired t-test shows that ID-6: t(24) = -2.921, p-value = 0.0077. Data shown as * p < 0.05/6 = 0.0083}
      \label{fig:result3.1}
   \end{figure}

\begin{figure}[ht]
      \centering
      \includegraphics[width=0.6\textwidth]{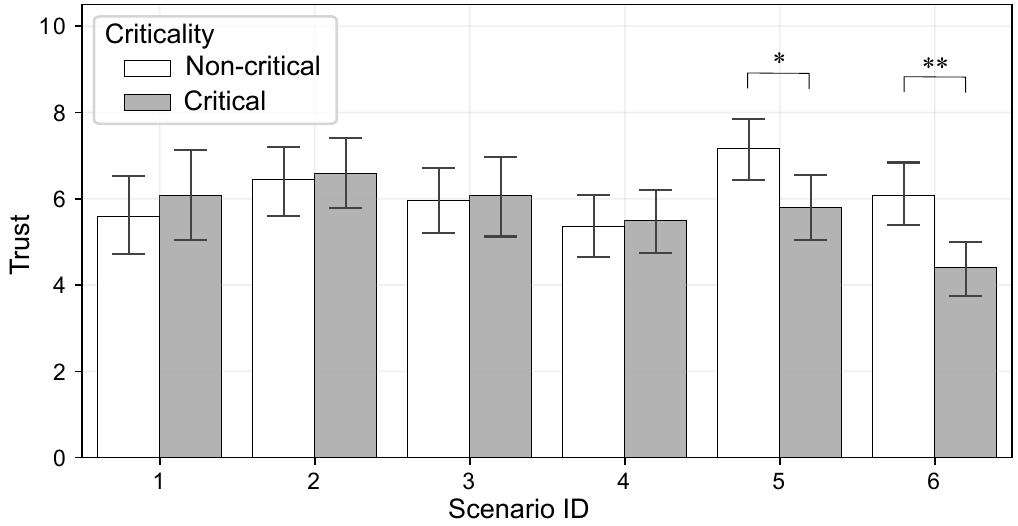}
      \caption{The trust score (H3.2) according to the criticality of the situation for each scenario ID (mean and 95\% confidence interval). The result of the paired t-test shows that ID-5: t(24) = -3.364, p-value = 0.0027 and ID-6: t(24) = -3.819, p-value = 0.0009. Data shown as * p < 0.05/6 = 0.0083, ** p < 0.01/6 = 0.0017}
      \label{fig:result3.2}
   \end{figure}

The results and discussion according to the third hypothesis are presented in this section. Fig.~\ref{fig:result3.1} and Fig.~\ref{fig:result3.2} show graphs of the reasonableness and trust scores according to the criticality for each scenario ID. It highlights that, in the critical situation (100~\unit{m}), the reasonableness score in the scenario ID-6 and the trust scores in the scenarios of ID-5 and ID-6 are significantly lower than in the non-critical situation (200~\unit{m}), although no significant differences were observed in the other IDs. Notably, the weak deceleration scenarios (ID-5 and ID-6) require a longer time to complete the preparation behaviour compared to the others. The significance level has been adjusted using the Bonferroni correction. %to avoid the multiple testing problem.

Hence, particularly in a time-sensitive situation, the temporal aspect of completing the merging is crucial in establishing trust with the host vehicle. This implies that, contrary to the third hypothesis, the host vehicle did not exhibit higher tolerance for insufficient merging preparation, i.e.\ defective longitudinal positions or strong deceleration. Instead, it exhibits a higher sensitivity to lagging movements of the merging vehicle while cultivating trust. These observations align with the previous study, emphasizing the expectation for merging vehicles to execute their manoeuvres efficiently in urgent scenarios~\cite{Potzy2}.

\vspace{3mm}

In summary, for optimal negotiation in congested merging, it is recommended to: identify an adequate merging space where the merging vehicle can be positioned 
between the leading vehicle and the host vehicle (particularly in the middle or closer to the leading vehicle) while synchronizing the vehicle's speed; activate the turn indicator once the positioning is accomplished;
%closer to the leading vehicle than to the host vehicle or in the middle but not further back than half the distance between the host and the leading vehicle; synchronize the vehicle's speed and activate the turn indicator when effective positioning is accomplished
decelerate courteously concerning the host vehicle but without unnecessary delay, while ensuring an appropriate safety gap to the leading vehicle; finally, execute a lane change if there is enough space to merge. In critical situations, it is important to minimise the duration of the merging preparation, positioning the merging vehicle in the middle, and executing the merging as soon as the host vehicle creates enough space.

\vspace{3mm}

\begin{figure}[ht]
      \centering
      \includegraphics[width=0.5\textwidth]{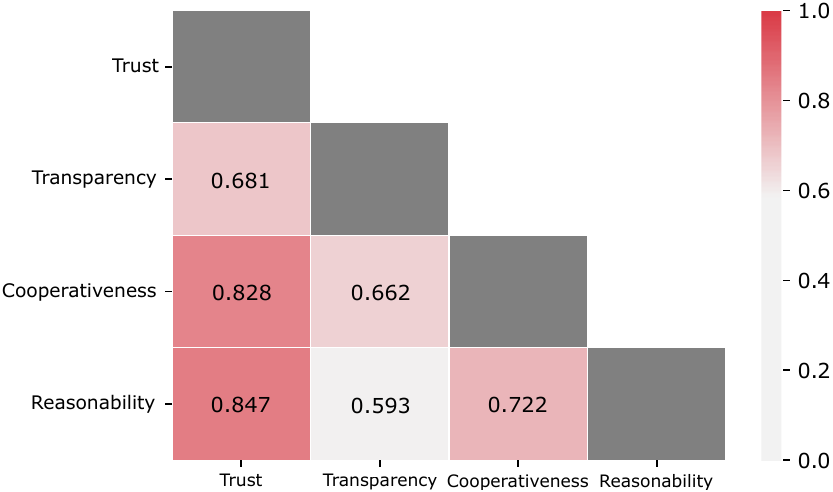}
      \caption{Correlation analysis of subjective items. The values are Pearson correlation coefficients.}
      \label{fig:correlation}
   \end{figure}

Finally, we conducted a multiple correlation analysis to examine the correlation between the subjective evaluation criteria for each hypothesis as illustrated in Fig.~\ref{fig:correlation}. The result demonstrates that cooperativeness and reasonableness are strongly correlated with trust, while transparency has a moderate correlation. Besides the results of previous studies that cooperative behaviour influences perceived trust~\cite{Zimmermann}, the host vehicle's driver considers the rationality of the behaviour regarding the urgency of the situation as a significant factor in trust formation. This partly aligns with the previous result that the drivers consider situational criticality as an influential factor in perceived cooperativeness~\cite{Stoll}.

\section{LIMITATION AND FUTURE WORK}

There are two limitations to this study that should be addressed in future work. Firstly, since the findings of this study are limited to the virtual environment, physical experiments will be necessary to validate the findings and to fully understand real-world driving behaviour. Because of the physical risk of being close to each vehicle and the difficulty of maintaining repeatability across the scenarios with various behavioural parameters, we adopted the simulation as the first step. However, actual vehicle testing should be conducted considering the difference in motion feedback and perceived criticality between virtual and real environments.

Secondly, this study conducted a subjective evaluation with participants possessing driving experience mainly in the United Kingdom. However, it is important to acknowledge the presence of distinct driving rules, styles and conventions across various countries or traffic cultures. Therefore, in order to ascertain the generalizability of the findings obtained from this study, a comprehensive examination of these cultural differences is imperative in future work.

\section{CONCLUSION}
Despite the limitations, this study provides valuable insights for designing a controller for AVs to operate in a merging scenario that requires mutual negotiation with human-driven vehicles. Specifically, the findings suggest that clear longitudinal positioning could improve the chance of successful merging, and cooperative deceleration within the merging lane as well as decisive actions in time-sensitive situations could enhance the level of trust perceived by the host vehicle in the congested merging scenario. These results could be leveraged to develop collaborative AVs that improve both safety and efficiency in real-world traffic situations.

\section*{ACKNOWLEDGMENT}

We would like to thank all the participants who kindly spared their valuable time for the experiment.

\bibliographystyle{unsrtnat}
%\bibliography{references}  %%% Uncomment this line and comment out the ``thebibliography'' section below to use the external .bib file (using bibtex) .


\end{document}